\documentclass[aps,prl,twocolumn,showpacs,superscriptaddress,groupedaddress]{revtex4}  
\usepackage{graphicx}  
\usepackage{dcolumn}   
\usepackage{bm}        
\usepackage{amssymb}   
\usepackage{amsmath}
\usepackage{color}

\def\be{\begin{equation}}
\def\ee{\end{equation}}
\def\bea{\begin{eqnarray}}
\def\eea{\end{eqnarray}}
\def\MeV{\mathrm{MeV}}

\begin{document}

\widetext


\title{A Theory of $X$ and  $Z$ Multiquark  Resonances}
\author{Luciano Maiani}
\email{Luciano.Maiani@cern.ch}
\affiliation{Theory Department, CERN, Geneva, Switzerland.
}
\author{Antonio D. Polosa}
\email{antonio.polosa@roma1.infn.it}
\affiliation{Dipartimento di Fisica and INFN,  Sapienza  Universit\`a di Roma, Piazzale Aldo Moro 2, I-00185 Roma, Italy.}
\author{Veronica Riquer}
\email{veronica.riquer@cern.ch}
\affiliation{Theory Department, CERN,  Geneva, Switzerland.}
\date{\today}

\begin{abstract}
We introduce the hypothesis that diquarks and antidiquarks in tetraquarks are separated by a potential barrier. We show that this notion can answer satisfactorily long standing questions challenging the diquark-antidiquark model of exotic resonances. The tetraquark description of $X$ and $Z$ resonances is shown to be compatible with present limits on the non-observation of charged partners $X^{\pm}$, of the $X(3872)$  and the absence of a hyperfine splitting between two different neutral states. In the same picture, $Z_c$ and $Z_b$ particles are expected to form complete isospin triplets plus singlets. It is also explained why the decay rate into final states including quarkonia are suppressed with respect to those having open charm/beauty states.
\end{abstract}

\maketitle

{\bf \emph{Introduction.}}  The observed lowest lying $X$ and $Z$ states  are found very close or slightly above the meson-meson thresholds with  the corresponding quantum numbers.  The $X(3872)$, $Z_c(3900)$, $Z_c^\prime(4020)$, $Z_b(10610)$, $Z_b^\prime(10650)$ axial resonances, have central mass values distant by  
\be
\delta=0\pm 0.195,\,+7.8,\, +6.7,\, +2.7,\,+1.8~\MeV
\label{uno}
\ee
 from the closer meson-meson thresholds with $1^+$ quantum numbers 
 \be
 {\bar D}^0 D^{*0},\,  {\bar D}^0 D^{*+},\, \bar D^{*0} D^{*+},\, \bar B^{0}B^{*+},\, \bar B^{*0}B^{*+} 
 \label{due}
 \ee 
 
Some authors believe that, being the $\delta$s  fairly small, different parametrizations of the lineshapes, combined with updated data analyses, might eventually show that $X,Z$ states  have masses below the aforementioned thresholds (see reviews~\citep{Ali:2017jda, Esposito:2016noz,  Chen:2016qju, Guo:2017jvc,Lebed:2016hpi,Olsen:2017bmm} and~\cite{Tornqvist:2004qy}). In the latter case the hadron molecule interpretation would become tenable, at least from the energetic point of view.  

With positive and finite $\delta$ values, a reasonable alternative description is in terms of compact tetraquarks, as in the diquark-antidiquark  model~\cite{Maiani:2004vq,Maiani:2014aja}.
~The model can describe all observed exotic hadrons in a unique scheme, including cases like $Z(4430)$~\cite{Maiani:2007wz, Maiani:2008zz}, the $J/\psi\,\phi$ resonances~\cite{Maiani:2016wlq} and the heavier, positive parity, pentaquark ${\cal P}(4570)$ \citep{Maiani:2015vwa,Maiani:2015iaa}, which are problematic to fit in the molecular picture.  We have to underscore that the existence of exotic charged charmed resonances with decays into $\psi(nS)\,\pi^\pm, \rho^\pm\cdots$ was a prediction of the diquark-antidiquark model~\cite{Maiani:2004vq} and an unwanted/unnecessary feature for molecular models.

Four quarks produced in high-energy hadron collisions, or in $B$ meson decays, have different alternatives for clustering in color neutral states namely, the diquark-antidiquark alternative
\begin{equation}
\Psi_{\cal D}=(\epsilon_{ijk}\,Q^j q^k)\,(\epsilon^{i m n}\,\bar Q_m\bar q_n^\prime)=[Qq][\bar Q\bar q^\prime]
\label{psid0}
\end{equation} 
or the meson-meson alternatives 
\begin{equation}
\Psi_{\cal M}=(Q^i\bar q_i)\, (\bar Q_k q^{\prime k})~\,\,\,\text{or}\,\,\,~(Q^i\bar Q_i)\, (\bar q_k q^{\prime k})
\label{molse}
\end{equation}

The $\Psi_{\cal M}$ component is supposed to be in the continuum spectrum of a shallow potential with no bound states | a residual strong interaction tail at large distances. The $\Psi_{\cal D}$ component is instead a stationary state in the color binding potential. 

The mass of the tetraquark can be slightly higher than the sum of the masses of the two open charm singlets, because strong attraction in color singlet channels is stronger than in color anti-triplet channels. Thus, it is not surprising that the observed tetraquarks appear near to the corresponding meson thresholds, albeit being heavier.

If the recoil energy $E_0$ in the center of mass of the color singlets in $\Psi_{\cal M}$  is high enough~\cite{Bignamini:2009sk}, a pair of free mesons will   be detected. 
If $E_0$ is sufficiently low (a rare circumstance in prompt production from high energy hadron collisions) the color singlets might rescatter forming a  tetraquark state that decays back into a meson meson pair~\cite{Esposito:2016noz}. The diquark-antidiquark tetraquark can as well be produced promptly.

The fact that $E_0$ tends to be large in high-$p_T$ events in hadronic collisions at the LHC is compatible with the non-observation of loosely bound molecules, like deuteron, produced promptly in 
such kinematic conditions~\cite{Esposito:2015fsa}. On the other hand, the large prompt production cross section of $X(3872)$ at the LHC appears to be in contradiction with a loosely bound molecule interpretation~\cite{Bignamini:2009sk, Esposito:2013ada, Guerrieri:2014gfa}.  
  
\begin{figure}[h!]
\begin{center}
\includegraphics[width=8truecm]{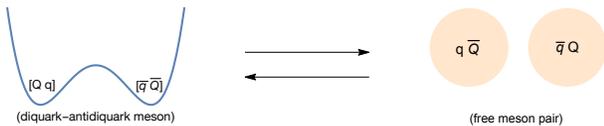}
\end{center}
\caption{Tunneling of  {\it light} quarks rearranges the diquark-antidiquark state $\Psi_{\cal D}$ (left panel) into two color singlets $\Psi_{\cal M}$ (right panel). The opposite process might proceed if the recoil energy between the color singlets is low enough to keep them in a small volume of configuration space.}
\label{f1}
\end{figure}

Following an argument of Selem and Wilczek~\cite{Selem:2006nd}, we make the hypothesis that a tetraquark can plausibly be represented by two diquarks in a double well potential separated by a barrier, as in Fig~\ref{f1}.
 
The argument can be summarised as follows. At large distances, diquarks see each other as QCD point charges and QCD confining forces are the same as in a quark-antiquark system.  At shorter distances, however, forces among different parts that tend to destroy the diquark, {\it e.g.} attraction between quarks and antiquarks, reduce the binding energy of the diquark. These effects  increase at decreasing distance and produce a repulsion among diquark and antidiquark~\cite{Selem:2006nd}, {\it i.e.} a component in the potential increasing at decreasing distance. If this effect wins against the decrease due to the color attraction, it will produce the barrier depicted in the figure. 

It is an hypothesis that we cannot prove, at the moment. However, it has some phenomenological support in the spectrum of $X(3872)$, $Z_c(3900)$, $Z_c^\prime(4020)$. Mass ordering indicates~\cite{Maiani:2014aja} 
that $i)$ spin-spin interactions between constituents located one  in the diquark and the other in the antidiquark  are definitely smaller than one would guess from the same interactions within mesons and $ii)$ the spin-spin interaction inside the diquark is about four times larger than the same interaction in the diquarks inside charmed baryon states. Thus the overlap probability $|\psi_{q{\bar q}^\prime}(0)|^2$ of a quark and an antiquark is suppressed and that of a quark pair $|\psi_{cq}(0)|^2$ is enhanced  with respect to what happens in mesons and  baryons respectively.

Fig.~\ref{f1}, taken literally, implies the existence of two length scales: the diquark radius, $R_{Qq}$ and the tetraquark radius, $R_{4q}$, which we assume to be well separated 
\be
\lambda=R_{4q}/R_{Qq}\geq 3
\label{ratio}
\ee 
In principle the diquark radius $R_{Qq}$ can be different if the diquark  is in a tetraquark or in a baryon. We will distinguish the latter naming it $R_{Qq}^{\rm baryon}$. 

Using established Constituent Quark Model techniques~\citep{CQM}, see also~\citep{Maiani:2004vq,Karliner:2016zzc}, we show that this picture can give a novel answer to the present lack of observation, in $B^{0,+}$ decays, of a second neutral state in the vicinity of the $X(3872)$ and of the associated charged state. We find that: $i)$ the two neutral states are quasi-degenerate within the mass resolution with which the $X(3872)$ is observed  and $ii)$ the associated charged state is produced much below the rate expected for a  pure isospin $I=1$ $X(3872)$ multiplet, complying with present limits. For the large charm quark mass, the two-lenghts picture leads, in addition, to $iii)$ an exponentially suppressed amplitude for  $X(3872)\to J/\psi~\pi\pi$, with respect to ${\bar D}^0 D^{* 0}$, qualitatively explaining the large branching fraction of the latter to the former mode, in spite of its much smaller phase space,  as observed in the phenomenology~\cite{Esposito:2016itg}. This behavior, as shown in~\cite{Esposito:2016itg}, is quite evidently shared by $Z_{c,b}^{(\prime)}$ resonances | the $Z(4430)$, being most likely a radial excitation, may have  slightly different features. 

An increase of the experimental resolution and statistics are crucial to support or disprove our picture, by searching for a double structure inside the $X(3872)$ line and for $X^\pm$ in the decays of $B$ mesons at lower branching fractions than at present. 

The $X^{\pm}$ charged resonances could also be produced prompt in proton-proton collisions at the LHC.
For the time being the prompt production of $X^0$ is well studied but no signs neither of  $X^\pm$  nor of  $Z_{c,b}^{\pm}$ are found. The experimental situation of $Z_c$ particles in $B$ decays is also unclear.

{\bf \emph{Isospin breaking in tetraquarks.}}
We recall the definitions
\bea 
&& X_u=\frac{1}{\sqrt{2}}\Big([cu]_0[{\bar c\bar u}]_1+[cu]_1[{\bar c\bar u}]_0\Big)\label{xup}\\
&& X_d=\frac{1}{\sqrt{2}}\Big([cd]_0[{\bar c\bar d}]_1+[cd]_1[{\bar c\bar d}]_0\Big)\label{xudown}
\eea
in brackets (anti)diquarks with the indicated flavors, in color ($\bm 3$) ${\bar {\bm 3}}$ and total spin indicated by the subscripts.

In~\cite{Maiani:2004vq,Maiani:2007vr}, we considered the mass difference $\Delta M=M(X_u)-M(X_d)$ to be determined by the $down$-$up$ quark mass difference 
\be
\Delta M=2(m_u-m_d)\approx -6~\mathrm{MeV} 
\ee

A more refined analysis~\citep{Rosner:1998zc,Karliner:2017gml} introduces the effect of Coulomb and hyperfine electromagnetic interactions and of the $u-d$ mass difference in the  strong hyperfine interaction. These effects are parametrised, for baryons, with three phenomenological parameters $a,\kappa,\gamma$ defined according to~\footnote{In the following equation write $m_{u}=\overline{m}+(m_u-m_d)/2$ and $m_{d}=\overline{m}-(m_u-m_d)/2$ where $\overline{m}=(m_u+m_d)/2$. Neglect $(m_u-m_d)^2/4$.  The coupling $g_s^2/\overline{m}m_c$ has to be rescaled by $\kappa_{cq}/\kappa_{cq}^{\rm Baryon}$ (where $\kappa_{cq}\equiv \kappa_{cq}^{\rm diquark} $) and it is used $\kappa_{cq}^{\rm Baryon}=g_s^2/\overline{m}m_c\, |\psi_B(0)|^2$.}
\bea 
&&{\rm \underline{Electrostatic}}\nonumber \\
 H_{ij}&=& Q_i\, Q_j~a\times\left( \frac{R_{Qq}^{\rm baryon}}{R_{ij}}\right) \label{elec}\\
&&{\rm \underline{Electromagnetic~hyperfine}}\nonumber \\
H_{q,c}&=&(Q_u-Q_d) Q_c\frac{\alpha}{\overline{m} m_c}{\bm S}_{q}{\bm \cdot}{\bm S}_c~|\psi(0)|^2=\nonumber \\
&=&2\gamma\,(Q_u-Q_d) Q_c\frac{\overline{m}}{m_c} \frac{|\psi(0)|^2}{|\psi_B(0)|^2}~2{\bm S}_q{\bm \cdot}{\bm S}_c \label{hfem}\\
&&{\rm \underline{Strong~hyperfine}}\nonumber \\
 \Delta H_{q,c}&=&\frac{g_s^2}{m_c} \left(\frac{1}{m_u} - \frac{1}{m_d} \right){\bm S}_{q}{\bm \cdot}{\bm S}_c \,|\psi(0)|^2=\nonumber \\
&=&-\kappa_{qc}\frac{m_u-m_d}{ \overline{m}} \frac{|\psi(0)|^2}{|\psi_B(0)|^2}~2{\bm S}_q{\bm \cdot}{\bm S}_c \label{hfstrong}
\eea
where we indicate explicitly the dependence from $m_{u/d}$ and $\overline{m}$ denotes the average light quark mass and a sum of the two charge conjugate contributions is understood. $R_{ij}$ can be either $R_{Qq}$ or $R_{4q}$, Eq.~(\ref{ratio}), and $R_{Qq}^{\rm baryon}$ is the radius of the diquark in the baryon.  $|\psi(0)|^2$ and $|\psi_B(0)|^2$ represent the $cq$ overlap probabilities in tetraquarks and baryons respectively. 
 
With the definitions in Eqs.~(\ref{elec}) to (\ref{hfstrong}) and defining $\Delta_m= m_u-m_d$, one finds the mass differences
\bea
&&M(X_u)-M(X_d)=\nonumber \\
&&=2\Delta_m+\frac{4}{3}a^\prime-\frac{5}{3}\frac{a^\prime}{\lambda}+\kappa_{cq}^\prime \frac{\Delta_m}{\overline{m}}-\frac{4}{3} \gamma^\prime \frac{\overline{m}}{m_c}\label{ndiff}\\
&&M(X_u)-M(X^+)=\nonumber \\
&&=\Delta_m+\frac{2}{3}a'-\frac{4}{3}\frac{a^\prime}{\lambda}+\kappa_{cq}^\prime \frac{\Delta_m}{2 \overline{m}}-\frac{2}{3} \gamma^\prime \frac{\overline{m}}{m_c}\label{nch}
\eea
Primed quantities refer to (anti)diquarks in tetraquarks and have to be scaled using the ratio of the hyperfine strong couplings, $\kappa_{cq}$ and $\kappa^\prime_{cq}$ in baryons and tetraquarks. 
The term $a'/\lambda$, representing the electrostatic attraction between diquark and antidiquark, has been further rescaled to the tetraquark radius.
We find $\kappa^\prime_{cq}=67$~MeV, from the mass difference of $Z(4020)$ and $Z(3900)$~\cite{Maiani:2004vq} and $\kappa_{cq}=15$~MeV, from the hyperfine mass differences of single charm baryons~\citep{Alibook}. Ref.~\citep{,Karliner:2017gml} finds  $\kappa_{cq}=19$~MeV. We take $\kappa_{cq}=17\pm 2$~MeV as an indication of the error. Accordingly, 
\be
 r=\frac{\kappa'_{cq}}{\kappa_{cq}}=3.94\pm0.45,~\frac{R_{cq}^{\rm baryon}}{R_{cq}}=r^{1/3}=1.58\pm 0.06\label{scales}
\ee

From a fit to the isospin violating mass differences of light baryons, Ref.~\citep{Karliner:2017gml} obtains: $2\Delta_m=-4.96;~a=2.83;~\gamma=-1.30,~\overline{m}=308,~m_c=1665$. Thus we obtain:  $a'=4.47;~\gamma' \overline{m}/m_c=-0.95$  (all in MeV). Numerical results are shown in Tab.~\ref{numerics}.
\begin{table}[htb!]
\centering
    \begin{tabular}{||c|c|c||} \hline
-- &{\footnotesize$\lambda=1 $}& {\footnotesize $\lambda=3$} \\ \hline
{\footnotesize$M(X_u)-M(X_d)$}& {\footnotesize $-6.1\pm0.1$}  & {\footnotesize $-1.2\pm 0.3$}\\
{\footnotesize$M(X_u)-M(X^+) $}&{\footnotesize  $-5.31\pm 0.05 $} &{\footnotesize  $ -1.34\pm 0.12$}\\
\hline
\end{tabular}
 \caption{\footnotesize Numerics of mass differences, in MeV, vs $\lambda$ in Eq.~(\ref{ratio}).}
\label{numerics}
\end{table}

The separation of the two scales makes a big effect. For $\lambda=1$, the electrostatic repulsion in the (anti)diquark is almost compensated by the electrostatic attraction between the diquark and the antidiquark, and the mass difference is dominated by $\Delta_m$. As we get to $\lambda=3$, the electrostatic repulsion dominates and the mass difference is greatly reduced, to the extent that $X_{u,d}$ may be considered to be quasi-degenerate, within the present experimental resolution of about $1$~MeV. The result justifies why only one line is seen in the $D^0 {\bar D}^{*0}$ channel and none in $D^+ D^{*-}+D^- D^{*+}$.  
$X^+$ is expected to be below threshold for the decay into $D^0 D^{*+}+D^+ D^{*0}$ but  it should be found among the products of charmonium decays of $B$ mesons, however within the bounds we shall consider now. 
 
{\bf \emph{Charmonium decays of $\bm B$ mesons.}} 
Starting from the overall weak process with one $q\bar q$ pair from the sea:
\be
[{\bar b} d]_{B^0} \to {\bar c} ~c {\bar s}+(d {\bar d}, {\rm or}~u{\bar u})+d \nonumber
\ee
one can describe the decays $B\to X~K$ with two amplitudes, corresponding to the kaon being formed from the $\bar s$ with the spectator $d$ quark, $A_1$, or with a $d$ or $u$ quark from the sea,  $A_2$. 

In particular
\bea
&& \mathrm{Amp(}B^0\to X_d\, K^0)\sim A_1+A_2\notag\\
\label{unox}
&& \mathrm{Amp(}B^0\to X_u\, K^0)\sim  A_1\\
&& \mathrm{Amp(}B^0\to X^-\,K^+) \sim A_2\notag
\eea 
and
\bea
&& \mathrm{Amp(}B^+\to X_d\, K^+)\sim A_1\notag\\
\label{duex}
&& \mathrm{Amp(}B^+\to X_u\, K^+)\sim A_1+A_2\\
&& \mathrm{Amp(}B^+\to X^+\,K^0 ) \sim A_2\notag
\eea 

With near degeneracy of $X_{u,d}$, even a small $q\bar q$ annihilation amplitude inside the tetraquark could produce sizeable mixing. We consider the mass eigenstates in the isospin basis, namely
\bea
&&X_1=\cos \phi~ \frac{X_u+X_d}{\sqrt{2}}+\sin \phi~\frac{X_u-X_d}{\sqrt{2}}\nonumber \\
&&X_2=-\sin \phi~ \frac{X_u+X_d}{\sqrt{2}}+\cos \phi~\frac{X_u-X_d}{\sqrt{2}}\label{mixmass}
\eea
It is straightforward to compute the rate for $B$ going to $X(3872)$, the sum of two unresolved, almost degenerate lines, followed by decay into $J/\psi+2\pi/3\pi$, as function of $\phi$ and of the ratio of the isospin zero and one amplitudes, $2\alpha=2A_1+A_2$, $2\beta=A_2$, respectively. Note that, when going from $B^0$ to $B^+$ in the $3\pi$ to $2\pi$ ratio, $\alpha \to \alpha,~\beta\to-\beta$.

\begin{figure}[h!]
\begin{center}
\includegraphics[width=10truecm]{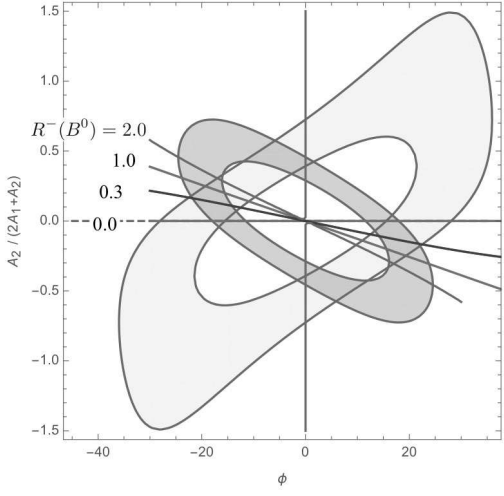}
\end{center}
\caption{Contour regions of $F^0(\phi,\frac{A_2}{2A_1+A_2})$, light shaded, and $F^+(\phi,\frac{A_2}{2A_1+A_2})$, shaded, see text. Four overlap areas correspond to regions of parameters which reproduce the experimental values of both $F^+$ and $F^0$. Solutions close to $\phi=0$ correspond to $R^-(B^0)\sim 2$ and are not acceptable. Solutions close to $\phi\sim\pm 20^0$ correspond to $R^-(B^0)\leq 2$. As indicated by level curves reported in the figure, a good fraction of the allowed region is compatible with the present limit $R^-(B^0)< 1$, see~\citep{Patrignani:2016xqp}, and with $R^+(B^+)<0.5$ (not reported in the figure). The center of the allowed region corresponds to 
$R^-(B^0)=0.3$ and $R^+(B^+)=0.2$.
}
\label{param}
\end{figure}

From PDG~\citep{Patrignani:2016xqp} we find close values of the two ratios within errors
\bea
R(B^0)&=&\frac{\Gamma (B^0\to K^0~X(3872)\to K^0\, J/\psi\, 3\pi)}{\Gamma (B^0\to K^0~X(3872)\to K^0\, J/\psi\,2\pi)}\nonumber \\
&=&1.4\pm 0.6=\frac{p_\rho}{p_\omega}~F^0\left(\phi,\frac{\beta}{\alpha}\right)\\
R(B^+)&=&\frac{\Gamma (B^+\to K^+\, X(3872)\to K^+ \,J/\psi\,3\pi)}{\Gamma (B^+\to K^+ ~X(3872)\to K^+\,J/\psi\, 2\pi)}\nonumber \\
&=&0.7\pm 0.4=\frac{p_\rho}{p_\omega}~F^+\left(\phi,\frac{\beta}{\alpha}\right)
\eea
where $p_{\rho,\omega}$ are decay momenta (averaged over Breit-Wigner distributions, see~\citep{Maiani:2004vq}). Fig.~\ref{param} reports the contour plots of the two experimental ratios $R(B^{+,0})$. We also define
\bea
R^-(B^{0})&=&\frac{\Gamma (B^{0}\to  K^+ X^- \to K^+ \,J/\psi\, \rho^-)}{\Gamma(B^0 \to  K^0 X(3872)\to K^0\, J/\psi\, \rho^0)}\nonumber \\
&=&G^-\left(\phi,\frac{\beta}{\alpha}\right)\\
R^+(B^+)&=&G^+\left(\phi,\frac{\beta}{\alpha}\right)=G^-\left(\phi,-\frac{\beta}{\alpha}\right) 
\eea
The two allowed regions with $\phi\sim \pm 20^0$ are compatible with the present limits $R^-(B^0), R^+(B^+)< 1$, see~\citep{Patrignani:2016xqp}. The center of the allowed region corresponds to $R^-(B^0)=0.3$ and $R^+(B^+)=0.2$.

{\bf \emph{Tunneling.}} The diquark-antidiquark system can rearrange itself into a color singlet pair of the type $\Psi_{\cal M}$ by exchanging quarks  through a tunneling transition. 

The small overlap between the constituent quarks in different wells suppresses quark-antiquark direct annihilation even in neutral tetraquarks and it leaves us with a two stage process: $i)$ switch of a quark and an antiquark among the two wells $ii)$ evolution of the quark-antiquark pairs (in their colour singlet component) into the corresponding mesons.

To illustrate the structure of decay amplitudes, we consider the state made by a diquark 
 localized at $x$ and an antidiquark localized at  $y$, 
 ~with $u$ and $\bar u$ light quarks as in  
\be
\Psi_{\cal D}\, =[cu](x)[\bar c\bar u](y)
\label{icsu1}
\ee
We can cluster quarks and antiquarks together by a Fierz rearrangement on color indices, which leads to, {\it e.g.}  
\be
\Psi_{\cal D}\, \sim \left(c(x)\bar  u(y)\right)\left(\bar c(y) u(x)\right)
\ee
(round brackets indicate that we have to take the projections over colour singlets). However this is not enough, since we still need to bring the light quark and the antiquark in the respective positions of $\bar c$ and $c$ ($y\leftrightarrow x$). This involves tunneling below the barrier between the two wells, Fig.~\ref{f1}.
The amplitude for a heavy quark tunneling is exponentially suppressed  with the mass of the heavy quark $\sim\exp(-\sqrt{m_c E}~\ell)$, where $E$ and ${\ell}$ are height and the extension of the barrier, so that: \textit{compact tetraquark couplings are expected to favour the open charm/beauty modes with respect to  charmonium/bottomonium ones}.

In addition, tunneling may provide dynamical factors in front of the various components of the Fierz rearranged expression. Including the diquark spins (subscripts), consider the states
\bea
&&\Psi^{(1)}_{\cal D}=[cu]_0[{\bar c\bar u}]_1\notag\\
&&\Psi^{(2)}_{\cal D}={\cal C}\Psi^{(1)}_{\cal D} =[cu]_1[{\bar c\bar u}]_0
\eea
with ${\cal C}$ the charge conjugation operation. We start by performing a Fierz rearrangement on color indices of $\Psi^{(1)}_{\cal D}$ and focus on the first (leading)  term 
\be
\Psi^{(1)}_{\cal D}\sim [c^\alpha \sigma_2 u^\beta](x)[{\bar c}_\beta\sigma_2{\bm \sigma}{\bar u}_\alpha](y)
\ee
which encodes the $c\bar u$ and $u\bar c$ color singlets (and singles out $c\bar c$ terms). After a Fierz rearrangement of spin indices we get
\bea
\Psi^{(1)}_{\cal D}&=&A[c^\alpha(x) \sigma_2 {\bar u}_\alpha(x)][{\bar c}_\beta(y)\sigma_2{\bm \sigma}u^\beta(y)] \nonumber \\
&-&B[c^\alpha(x) \sigma_2{\bm \sigma} {\bar u}_\alpha(x)][{\bar c}_\beta(y)\sigma_2 u^\beta(y)]+\\
&+& iC  [c^\alpha(x) \sigma_2{\bm \sigma} {\bar u}_\alpha(x)]{\bm \times} [{\bar c}_\beta (y)\sigma_2{\bm \sigma} u^\beta(y)]\notag
\eea
$A,~B,~C$ are non-perturbative coefficients  associated to  different barrier penetration amplitudes for {\it different light quark spin configurations}. Using an evident meson field notation we can write
\be
\Psi^{(1)}_{\cal D}=A\,D^0 {\bar {\bm D}}^{*0}- B\,{\bm D}^{*0}{\bar D}^0+iC\,{\bm D}^{*0}{\bm \times} {\bar {\bm D}}^{*0}
\ee
Similarly 
\be
\Psi^{(2)}_{\cal D}=B\, D^0 {\bar {\bm D}}^{*0}-A\,{\bm D}^{*0}{\bar D}^0-iC\,{\bm D}^{*0}{\bm \times} {\bar {\bm D}}^{*0}
\ee

{\bf \emph{$\bm X_u$, $\bm X_d$ and $\bm X^{\pm}$.}} 
Following Eqs.~(\ref{xup}) and (\ref{xudown}), $X_u$ can be casted in the form
\be
 X_u\sim \frac{\Psi^{(1)}_{\cal D}+\Psi^{(2)}_{\cal D}}{\sqrt{2}}= \frac{A+B}{\sqrt{2}}\, (D^0 {\bar {\bm D}}^{*0}- {\bm D}^{*0}{\bar D}^0)
 \label{xuprime}
 \ee
whereas 
\be
 X_d\sim \frac{A+B}{\sqrt{2}}\, (D^+ {\bm D}^{*-}- {\bm D}^{*+} D^-)
 \label{xusec}
 \ee

Similar considerations apply to $X^{\pm}$, described by 
\be
X^\pm\sim  \frac{A+B}{\sqrt{2}}~(D^\pm {\bar {\bm D}}^{*0}- {\bm D}^{*\pm}{\bar D}^0)
\label{charged}
\ee
With the results of Tab.~\ref{numerics}, $X_d$ is below threshold for the decay suggested by (\ref{xusec}). Both mass eingenstates in (\ref{mixmass}) decay in  $D^0~{\bar D}^{0*}$ via mixing. Charged partners are also lighter than the corresponding meson thresholds in (\ref{charged}) and their decay occurs via the subleading charmonium decays considered below.

{\bf \emph{$\bm Z_c^{(\prime)}$ and  $\bm Z_b^{(\prime)}$.}} 
In the case of $Z_c$ and $Z_b$ resonances, charged and neutral states are observed. Two neutral tetraquarks  are expected in this case too, although potentially quasi-degenerate. 

Consider the neutral, $u\bar u$  component of the $Z_c$ multiplet
\bea
\label{zneutral}
Z_c&=&\frac{1}{\sqrt{2}}\Big([cu]_0[{\bar c\bar u}]_1-[cu]_1[{\bar c\bar u}]_0\Big)= \\
&=& \frac{A-B}{\sqrt{2}}\, (D^0 {\bar {\bm D}}^{0*}+ {\bm D}^{0*}{\bar D}^0) +i\sqrt{2} C\, {\bm D}^{*0}{\bm \times} {\bar {\bm D}}^{*0}\nonumber
\eea
The non-trivial dependence of tunneling factors from the light quark spin ({\it i.e.} $A\neq B$ unlike in the naive Fierz transformation), allows $Z_c$ to decay in  $D^0 \bar{\bm D}^{*0}$, the decay in $\bm D^{*0} \bar{\bm D}^{*0}$ being  forbidden by phase space. The $d\bar d$ component would be coupled to the neutral combination of charged charmed mesons. The two decay channels for the mass eigenstates might get mixed.  

The expression for charged states follows naturally from~\eqref{zneutral}, but this time, (see~\eqref{uno}), there is enough phase space to decay into charged open charm components. 

The $Z_c^\prime$ resonances are constructed in a very similar way, with different non-perturbative coefficients in~\eqref{zneutral}, {\it e.g.}
\bea
Z'_c&=&\Big([cu]_1~ [{\bar c\bar u}]_1\Big)_{J=1}=\nonumber \\
&=& E~ (D^0 {\bar {\bm D}}^{0*}+ {\bm D}^{0*}{\bar D}^0) +iF~{\bm D}^{*0}{\bm \times} {\bar {\bm D}}^{*0}
\label{zprimeneutral}
\eea

An interesting experimental check is that of studying the mass difference between the charged and neutral components of the $Z_c$ resonance, which we would expect to be almost degenerate, as is the case for the $X$.

 There are no qualitative differences in the description of the $Z_b$ and $Z_b^\prime$ resonances except the fact that thresholds are closer, as indicated in~\eqref{uno} -- this could be due to the reduced chromomagnetic couplings by the large $b$ quark mass. As a consequence, the analog of the $X(3872)$ in the beauty sector could be pushed below threshold by spin interactions and forced to decay in the subleading bottomonium modes.

{\bf \emph{Sub-leading decays.}}  Heavy quark tunnelings amplitudes do not vanish for finite heavy quark masses. In particular it is found
\bea
\label{xus}
X_u&\sim& a\,i{\bm J/\bm\psi}{\bm \times}({\bm \omega}^0+{\bm \rho}^0)\\
Z_u&\sim&b\,\eta_c ({\bm \omega}^0+{\bm \rho}^0) -c\,{\bm J/\bm\psi}(\eta_q+\pi^0)
\label{zus}
\eea
while 
 \be
 \label{zpus}
Z^\prime_u\sim d\,\eta_c ({\bm \omega}^0+{\bm \rho}^0) +e\, {\bm J/\bm \psi}(\eta_q+\pi^0)
\ee
where the non-perturbative coefficients $a,b,...,e$ are all equal in the limit of naive Fierz couplings. 
The formulae for $X_d,~Z_d,~Z^\prime_d$ are obtained by letting ${\bm \rho}^0 \to -{\bm \rho}^0$ and $\pi^0 \to -\pi^0$.

For an orientative estimate, we may take the leading semiclassical approximation of tunneling amplitudes (see~\cite{Landau:1977}) 
\be
{\cal A}_M\sim e^{-\sqrt{2ME}\ell}
\ee
We use the quark masses, $m_q$ and $m_c$, quoted before from Ref.~\citep{Karliner:2017gml}, the orientative values: $E=100$~MeV and $\ell=2$~fm 
to obtain, neglecting factors of $\cal O$(1)
\be
R=\left(\frac{a}{A+B}\right)^2\sim \left(\frac{{\cal A}_{m_c}}{{\cal A}_{m_q}}\right)^2\sim 10^{-3} 
\ee
With decay momenta (in MeV): $p_\rho\sim 124$~\citep{Maiani:2004vq}, $p_{D D^*}\sim 2$~\citep{Patrignani:2016xqp}, one would find
\be
\frac{\Gamma(X(3872)\to J/\psi~\rho)}{\Gamma(X(3872)\to D{\bar D}^*)}=\frac{p_\rho}{p_{D D^*}}~R\sim 0.1
\ee
compatible with: $B(X(3872)\to J/\psi~\rho)\sim 2.6\times 10^{-2}$, $B(X(3872)\to D{\bar D}^*) \sim 24\times10^{-2}$~\citep{Patrignani:2016xqp}.

{\bf \emph{Conclusions.}}   In this paper we have analyzed  the typical objections raised against the tetraquark model in the diquark-antidiquark realization. The  replies we provide are based on a picture of the diquark correlations in hadrons, that we have advocated several times in the past, and examined now in all of its consequences. On this basis we show that the neutral and charged components of $X$ could be quasi-degenerate. As a consequence, the $X^\pm$ should not be observed in open charm decays but only in final states containing charmonia. However  the charged $X$ may have much smaller branching fractions in $B$ meson decays than expected and this requires  some dedicated experimental  effort to go beyond the bounds which have been set years ago.  The decay modes of the $Z^{(\prime)}$ particles are also explained and their occurrence in isospin triplets is understood. A number of questions on the $Z_{c,b}$ particles are left open by the experiment | all of them have a crucial role to the assessment of the considerations made here. In particular all $X,Z$ resonances should be produced in prompt $pp$ collisions, whereas there are no hints yet on $Z$ particles in these production channels. Also, $Z$s should be seen in $B$ decays too and a similar hyperfine structure of neutral $Z$ could  eventually be resolved.

\end{document}